\documentclass{mem}
\usepackage{natbib}\usepackage{txfonts}\usepackage{balance}
\usepackage{graphicx}
\usepackage[a4paper]{hyperref}
\idline{1}{1}
\begin{document}
\def\teff{$T\rm_{eff }$}
\def\kms{$\mathrm {km s}^{-1}$}

\title{
Hydrodynamics of a perturbation in the jet of 3C~111
}

   \subtitle{}

\author{
M. Perucho\inst{1} \and
        I. Agudo \inst{2} \and
        J.~L. G\'omez \inst{2} \and
        M. Kadler\inst{3,4,5} \and
        E. Ros\inst{1} \and
        Y. Y. Kovalev\inst{1,6}
          }

  \offprints{M. Perucho, \email{perucho@mpifr-bonn.mpg.de}}

\institute{Max-Planck-Institut f\"ur Radioastronomie, Auf dem H\"ugel 69, D-53121 Bonn, Germany 
\and
Instituto de Astrof\'{\i}sica de Andaluc\'{\i}a (CSIC),  Apartado 3004, E-18080 Granada, Spain 
\and 
Dr. Karl Remeis-Observatory, University of Erlangen-Nuremberg, Sternwartstrasse 7, D-96049 Bamberg, Germany \and 
CRESST/NASA Goddard Space Flight Center, 662 Greenbelt, MD 20771, USA 
\and 
Universities Space Research Association, 10211 Wincopin Circle, Suite 500 Columbia, MD 21044, USA 
\and
Astro Space Center of Lebedev Physical Institute, Profsoyuznaya 84/32, 117997 Moscow, Russia}

\authorrunning{Perucho et al.}

\titlerunning{Hydrodynamics of a perturbation in the jet of 3C~111}

\abstract{We present results on the modelling of the ejection of a superluminal
component in the jet of \object{3C~111}. We propose that the component
is generated by an injection of dense material followed by a decrease in the injection rate of bulk particles in the jet. Our model is supported by 1D relativistic hydrodynamics and emission simulations, and is capable of reproducing the brightness evolution of two features, as revealed by 15 GHz VLBA observations. We show that other scenarios, such as an increase of the Lorentz factor in the material of the perturbation,
fails to reproduce the observed evolution of this flare.

}
\maketitle{}

\section{Introduction}
Flaring events at radio frequencies are known to take place in
Active Galactic Nuclei (AGN), usually followed by the observation
of new radio features in the parsec-scale jets \citep[e.g.,][]{sa02}. It has been shown that the ejection of those features, or components, is related to dips in the X-ray emission from the active nucleus in the case of 3C~120 \citep{Mar02}, and perhaps also in 3C~111 \citep{Mar06}. The dips in X-rays precede the observations of new radio-components. The decrease in X-ray emission may be caused by the loss of the inner regions of the disc. In this scenario, a fraction of the accreted material is injected in the jet and a new component is later observed in VLBI images,
after the material becomes detectable at the observing frequencies, as it evolves downstream. The components are interpreted as the shocks produced by the ejection of denser and/or faster plasma in the flaring event from the accretion disc \citep{mar85}. The conditions for triggering the ejection of the material in those radio features are still unknown. Hydrodynamical simulations \citep[][ A03 hereafter]{Alo03} have shown that such jet perturbations produce a forward and a reverse structure, which would be expected to be observed as a fast front and a slower back component.

\begin{figure}[t]
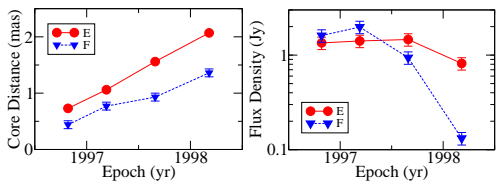

%\centering
 \includegraphics[clip,angle=0,width=0.23\textwidth]{fig1a}
 \includegraphics[clip,angle=0,width=0.24\textwidth]{fig1b}
 \caption{Core distance and flux density evolution with time of components E and F in 3C~111, based on the results from K08.}
%\includegraphics[clip,angle=0,width=0.24\textwidth]{fig1}
%\caption{Different epochs showing the evolution of components E and F in 3C~111, based on the %results from K08.}
\label{fig:0}
\end{figure}

In the jet of 3C~111 ($z=0.049$, $1\,\rm{mas}\simeq 1\,\rm{pc}$), a very strong flaring event in early 1996 gave rise to the ejection of two jet features observed at 15 GHz with the Very Long Baseline Array \citep[labeled as components E and F -see Fig.~\ref{fig:0} and][ K08 hereafter]{Kad08}. Both component trajectories can be back-extrapolated to similar ejection epochs within 3 months (around 1996.10). However, they show different speeds and the time
evolution of their brightness is different (see Fig.~\ref{fig:0}):
whereas the inner component F is initially brighter (1996.82 and 1997.19)
and fades out very rapidly (1997.66 and 1998.18), the leading component E
shows a slower decrease in flux density. After 1999,
F has disappeared and E evolves accelerating. In \cite{pe08} we propose the possibility that these components are the front and rear region of a single perturbation. Here we review this result and discuss on the problems that other possible scenarios could face. 

\section{RHD and Emission Simulations}
We have performed one-dimensional numerical relativistic hydrodynamics (RHD) simulations in
which a square perturbation in density is injected into a steady jet, without modifying the initial Lorentz factor, and relaxing the condition that the initial jet flow is reestablished immediately after the perturbation. We have substituted this by a rarefied flow, representing a reduction of the injection rate. In this picture, the original jet injection rates should be recovered after some time. However, in this work we only focus on the evolution of the strong ejection and the period before the reestablishment of the jet flow. Multidimensional simulations are out of the scope of this work due to the computational effort required and to the one-dimensional character of this problem. The simulations have been performed using a numerical code that solves the equations of relativistic hydrodynamics written in the conservation form, as described in \cite{pe05} and \cite{mart97}.

\begin{figure*}[!t]
   \centering
\includegraphics[clip,angle=0,width=\linewidth]{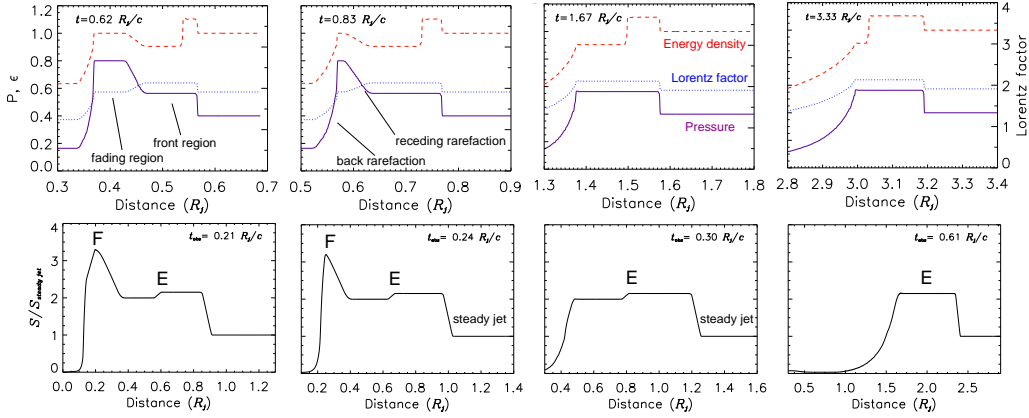}
\caption{
Snapshots of the evolution (left to right) of a square perturbation
injected in a steady jet, followed by a strong rarefaction. The
dotted-light-blue lines stand for Lorentz factor, the solid-dark-blue
line stands for pressure and the dashed-red lines stand for specific internal energy. 
The simulation is run with 24000~cells; the velocity of the initial flow is $v_j=0.9\,c$; the perturbation is
injected with a density twice that of the jet and velocity $v_p=0.9\,c$; the rarefied medium is injected after the perturbation with the same velocity as the initial flow, and pressure ten times
smaller than that of the initial flow. Please note the change of scale in the abcissae. The bottom panels show the simulated total intensity emission along the jet axis at four representative epochs. The identification of the features in the simulation with the observed
components E and F in K08 is indicated in each panel. A jet width of 100 cells and axial simmetry is used to compute the emission.}
\label{fig:1}
\end{figure*}

The details of the simulation are given in the caption of Fig.~\ref{fig:1}. The top panels in Fig.~\ref{fig:1} show different snapshots of the evolution of the
square perturbation injected in a steady flow, in pressure, Lorentz factor and specific internal energy.

Using the RHD simulations outlined above as input, we have computed the
corresponding 1D optically thin radio synchrotron emission as seen by an
observer with a line of sight at $19^{\circ}$ to the jet axis (K08). For these
computations, we used the numerical code and the procedure described in
\cite{go97} and references therein. This code takes into
account all the relevant relativistic effects, including the light
travel time delays.

In the simulation (see Fig.~\ref{fig:1}), the \textit{front} region includes the leading part of the perturbation and is identified with component E in K08, whereas we define the \textit{fading} region as the rear part of the perturbation and identify it with component F (see Fig.~\ref{fig:1}). The material in the front region, consisting of shocked material from the steady jet and rarefied material from the perturbation separated by a contact discontinuity, shows smaller values for the pressure, and some acceleration due the propagation in the lower pression steady jet fluid. The material in the fading region crosses the receding rarefaction that separates it from the front region (top panels in Fig.~\ref{fig:1}) and it is also ``eroded'' by the back rarefaction. Consequently, the front structure evolves increasing its size as the front shock incorporates material from the steady jet and the material from the fading region crosses the receding rarefaction. Thus, the front region consists of the forward shock structure of the perturbation (E in Fig.~\ref{fig:1}), and the fading region is formed by the remains of the perturbation that have not crossed the receding rarefaction (F). The synchrotron emissivity (bottom panel in Fig.~\ref{fig:1}) is governed by the jet pressure and hence the emission evolution is very similar to the pressure evolution of the RHD simulations. In the emission results, the front region (component E) propagates without much flux density evolution since injection. However, the fading structure (component F), which initially shows a notably larger flux density than component E, rapidly decreases in emission as the receding and back rarefactions erode it. The reverse shock (see A03) is neither relevant nor observationally significant in our simulations, as it propagates in a very rarefied medium. For that reason it is not shown in Fig.~\ref{fig:1}.

Notice that the Lorentz factor values in Fig.~\ref{fig:1} are those corresponding to the fluid. In contrast, VLBI observations provide us with pattern velocities. In the simulation, the velocity of the front shock is measured to be $v_{s}\sim 0.96\,c$ ($v_{\rm{E}}^{obs} \sim 3.5 c$), whereas that of the fading region is
$v_{r}\sim 0.87\,c$ ($v_{\rm{F}}^{obs} \sim 1.7 c$), both similar to those found in the observations (K08). We remark the fact that the velocity of the material in the fading region is faster than that of the receding rarefaction (cf. Fig.\ref{fig:1}), as expected from the explanation in the previous paragraph. We also point out that the dilute material shown in Fig.~\ref{fig:1} presents a modified velocity due to passage through the reverse shock.

A second simulation was performed for a faster perturbation, with Lorentz factor $\Gamma=3.6$, while keeping the rest of the parameters as in the previous simulation. The results \citep[see ][]{pe08} show that the front region of the perturbation is overpressured with respect to the rear region and, thus, the former is brighter than the latter, as shown by the emission simulations. This is in clear contradiction with the observations of the jet in 3C~111 (Fig.~\ref{fig:0} and K08). The difference is due to the stronger front shock produced in this case. It is also important to mention that the wave separating both regions is now a reverse shock, instead of the receding rarefaction shown in Fig.~\ref{fig:1}. This is a general result for fast perturbations. 

\section{Discussion}
In any of the scenarios mentioned above, the perturbed regions have enhanced emission with
respect to the underlying jet. However, only 1) an overpressured perturbation with the same Lorentz factor as the underlying flow avoids the front region to be brighter from the beginning, and 2)
a rarefaction behind the perturbation avoids the formation of a strong reverse shock behind the perturbation. With these restrictions, the second component fades out rapidly and then the first one dominates the emission, as observed for components E and F in 3C111.

We note that a denser medium injected after the perturbation would lead to the formation of a brighter feature behind component F, produced by the reverse shock formed between the end of the perturbation and the medium injected afterwards. As stated above, this shock appears also in our simulations but it has a negligible effect in terms of emission. Observational support for the inclusion of this tenuous material in the simulations can be found in the prominent emission gap following behind the E/F complex in 3C~111 (see K08). New ejection of emitting material is detected on the time scale of more than 2 years, corresponding to a gap width of up to 2\,mas in 1999 (cf. K08). We have exagerated the effect of the dilute medium in order to focus on the evolution of the perturbation alone. However, such a dilute medium (a factor 10 less dense than the steady jet) is not needed for the conclusions of this work to be valid: a denser medium, but underdense with respect to the steady jet, would produce a similar effect, with little emission contributed from the reverse shock formed. 

Thus, our result explains the evolution of these radio-components on the basis of the scenario explained above. From this work, we can derive a recipe for distinguishing between ejection events composed by denser material only or denser and/or faster material. On top of this, we want to remark that these results are based on the hypothesis that the injection of bulk particles in the jet is reduced after the injection of the perturbation, which could be related to the processes taking place in the accretion-disk/black-hole system, in the line of the results in \cite{Mar02}, who showed the existence of a relation between the processes taking place in the accretion-disk and the jet \citep[see][for discussion on this issue]{pe08}. 

At present we are following a new radio-flare in this source with denser sampling. The aim is to perform a deeper analysis of the early stages of evolution of the expected radio-components and to compare this evolution with the model presented here. 

\begin{acknowledgements}
MP acknowledges support from a postdoctoral fellowship of the
``Generalitat Valenciana'' (``Beca Postdoctoral
d'Excel$\cdot$l\`encia''). IA is 
supported by an I3P contract with the Spanish ``Consejo Superior 
de Investigaciones Cient\'{i}ficas''. MP, IA and JLG acknowledge support by
the Spanish ``Ministerio de Educaci\'on y Ciencia'' and the European Fund
for Regional Development through grants AYA2007-67627-C03-01 and AYA2007-67627-C03-03. 
MK has been supported by the NASA Postdoctoral Program at the Goddard Space Flight Center, administered by the Oak Ridge Associated Universities through a contract with NASA. YK is a research fellow of the
Alexander von Humboldt Foundation. We thank J-M Mart\'{i} and M.-A. Aloy for useful discussion and comments. 
\end{acknowledgements}

\bibliographystyle{aa}

\begin{thebibliography}{}
%
\bibitem[Aloy et al.(2003)]{Alo03} Aloy, M.-{\'A}., et al. 2003, ApJ, 585, L109 (A03)
%Mart{\'{\i}}, J.-M$^{\underline{\rm a}}$, G{\'o}mez, J.~L., Agudo, I., M{\"u}ller, E.,
%\& Ib{\'a}{\~n}ez, J.-M$^{\underline{\rm a}}$ 2003, ApJ, 585, L109
%
\bibitem[G\'omez et al.(1997)]{go97} G{\'o}mez, J.~L., et al. 1997, ApJ, 482, L33 
%Mart{\'{\i}}, J.-M$^{\underline{\rm a}}$, Marscher, A.-P., Ib{\'a}{\~n}ez, J.-M$^{\underline{\rm %a}}$, Alberdi, A. 1997, ApJ, 482, L33
%
\bibitem[Kadler et al.(2008)]{Kad08} Kadler, M., et al. 2008, ApJ, 680, 867 (K08)
%Ros, E., Perucho, M., et al. 2008, ApJ, 680, 867 (K08) 
%
\bibitem[Marscher \& Gear(1985)]{mar85} Marscher, A.P., Gear, W.K. 1985, ApJ, 298, 114 
%
\bibitem[Marscher et al.(2002)]{Mar02} Marscher, A.P., et al. 2002, Nature, 417, 625 
%Jorstad,S.G., G\'omez, J.~L., et al. 2002,
%Nature, 417, 625
%
\bibitem[Marscher(2006)]{Mar06} Marscher, A.P. 2006, Astron. Nachrichten, 327, 217
%
\bibitem[Mart\'{\i} et al.(1997)]{mart97} Mart\'{\i}, J.-M$^{\underline{\rm a}}$, et al. 1997,
ApJ, 479, 151 
%M\"uller, E., Font, J.A.,
%Ib\'a\~nez, J.M$^{\underline{\rm a}}$, and Marquina, A. 1997,
%ApJ, 479, 151
%
\bibitem[Perucho et al.(2005)]{pe05}
Perucho, M., et al. 2005, A\&A, 443, 863
%Mart\'{\i}, J.-M$^{\underline{\rm a}}$, Hanasz, M. 2005, A\&A, 443, 863
%
\bibitem[Perucho et al.(2008)]{pe08}
Perucho, M., et al. 2008, A\&A Letters, submitted
%Agudo, I., G\'omez, J.~L., Kadler, M., Ros, E., Kovalev, Y.~Y. 2008, A\&A Letters, submitted
%
\bibitem[Savolainen et al.(2002)]{sa02}
Savolainen, T., et al. 2002, A\&A, 394, 851
%Wilk, K., Valtaoja, E., Jorstad, S.G., Marscher, A.P.  2002, A\&A, 394, 851
%
\end{thebibliography}

\end{document}